\documentclass[journal]{IEEEtran}

\ifCLASSINFOpdf
\else
\fi

\usepackage{booktabs}     
\usepackage{cite}
\usepackage{indentfirst}
\usepackage{graphicx}
\usepackage{amssymb}
\usepackage{amsfonts}
\usepackage{amsmath}
\usepackage{cases}
\usepackage{algpseudocode}
\usepackage{subfigure}
\usepackage{mathrsfs}
\usepackage{makecell}
\usepackage{array}
\newcolumntype{L}[1]{>{\raggedright\arraybackslash}p{#1}}
\usepackage[linesnumbered,boxed,ruled,commentsnumbered]{algorithm2e}
\SetKwRepeat{Do}{do}{while}%
\usepackage[ruled]{algorithm2e}
\usepackage{amsthm}
\usepackage{multirow}
\usepackage{url}
\usepackage{amssymb}
\usepackage{color}
\usepackage{textcomp}
\usepackage{afterpage}

\newtheoremstyle{baitingstyle}{3pt}{3pt}{\itshape}{1.2em}{\itshape}{:}{0.5em}{{\thmname{#1}\thmnumber{ #2}\thmnote{ (#3)}}}
\theoremstyle{baitingstyle}

\theoremstyle{plain}

\theoremstyle{baitingstyle}

\theoremstyle{baitingstyle}

\theoremstyle{baitingstyle}

\theoremstyle{baitingstyle}

\theoremstyle{baitingstyle}

\newcommand*{\QEDA}{\hfill\ensuremath{\blacksquare}} 

\begin{document}
%
\title{A Collision-Aware Predictive Rerouting Framework \\ for Connected and Automated Vehicles}
\author{Tony Kinchen,~Ting~Bai,~\IEEEmembership{Member,~IEEE,}~and~Andreas A.~Malikopoulos,~\IEEEmembership{Senior Member,~IEEE}
\thanks{This work was supported in part by NSF (Grant Numbers: CNS-2401007, CMMI-234881, IIS-2415478), the National Natural Science Foundation of China (Grant Number: 62603450), and MathWorks.
}\vspace{1.5pt}

\thanks{Tony Kinchen is with the Systems Engineering Program, Cornell University, NY-14853 Ithaca, the United States (e-mail: \texttt{tjk238@cornell.edu}).}
\thanks{Ting Bai is with the School of Automation and Intelligent Sensing, Shanghai Jiao Tong University, 200240 Shanghai, China (e-mail: \texttt{tingbai@sjtu.edu.cn}).}
\thanks{Andreas A. Malikopoulos is with the Applied Mathematics, Systems Engineering, Mechanical Engineering, Electrical and Computer Engineering, and School of Civil \& Environmental Engineering, Cornell University, Ithaca, NY,
USA. (email: \texttt{amaliko@cornell.edu}).}
}

\markboth{IEEE Transactions on Intelligent Transportation Systems}%
{Kinchen \MakeLowercase{\textit{et al.}}: Collision-Aware Predictive Rerouting for CAVs}
\maketitle

\begin{abstract}
In this paper, we investigate the problem of mitigating non-recurrent congestion in urban transportation networks through predictive and adaptive routing for connected and automated vehicles (CAVs). We propose a collision-aware predictive rerouting framework that integrates microscopic SUMO simulation, DCRNN-based spatiotemporal traffic-state forecasting, collision-window lookahead, and multi-objective route scoring. The routing controller uses predicted congestion, live emission information, route-length penalties, and incident-window constraints to decide when CAVs deviate from their current routes. The completed simulation campaign across three urban network scenarios, five CAV penetration levels, three random seeds, and multiple benchmark rerouting strategies shows that the proposed full framework does not dominate all baselines in aggregate travel time, distance, or emissions. Instead, its principal advantage is operational stability: it maintains comparable network-level efficiency while substantially reducing rerouting frequency relative to active rerouting baselines. These results demonstrate both the promise and the tradeoffs of embedding graph-based prediction and collision awareness into real-time CAV rerouting, and they motivate further work on forecast calibration, uncertainty-aware prediction, and system-level fairness between CAV and non-CAV traffic.
\end{abstract}


\IEEEpeerreviewmaketitle
\section{Introduction}\label{Section I}
\IEEEPARstart{T}{he} continued growth of urban populations and the increasing complexity of transportation networks have intensified the challenges associated with traffic congestion, safety, and environmental sustainability. Non-recurrent congestion caused by incidents and collision events further exacerbates inefficiencies in urban mobility systems, often leading to significant travel delays, elevated emissions, and increased collision risk. These disruptions are particularly difficult to manage because they arise unpredictably and propagate dynamically through interconnected road networks. In parallel, the advancement of connected and automated vehicle (CAV) technologies has created new opportunities for intelligent traffic management through real-time data exchange, predictive control, and improved traffic-flow stability under connected and automated driving environments \cite{olia2016assessing,talebpour2016influence,11615763}. Leveraging these capabilities to proactively mitigate congestion and reduce exposure to hazardous traffic conditions represents a critical step toward resilient and sustainable urban transportation systems, particularly because connected and automated vehicle technologies have been shown to offer measurable safety benefits under appropriate deployment conditions \cite{wang2020crashes}. Furthermore, congestion and stop-and-go traffic have been shown to significantly increase fuel consumption and $\mathrm{CO}_2$ emissions, while the environmental effects of CAV operations can vary by road type, driving behavior, penetration rate, and traffic-flow stabilization effects, reinforcing the need for integrated solutions that jointly address operational efficiency and environmental impact \cite{barth2008real,stern2019quantifying,bandeira2021potential}. In addition, empirical studies have demonstrated the importance of accurately quantifying congestion through metrics such as the Travel Time Index for effective traffic management and planning \cite{erdelic2021congestion}.

Microscopic traffic simulation platforms have emerged as indispensable tools for evaluating advanced traffic management strategies under controlled yet realistic conditions. The Simulation of Urban Mobility (SUMO) provides a flexible and extensible framework for modeling large-scale road networks and enabling real-time control via external interfaces~\cite{Behrisch2011}. In addition, enhanced emission modeling capabilities within SUMO allow for detailed estimation of vehicle-level environmental impacts under varying traffic conditions \cite{Krajzewicz2014}. Recent research has demonstrated the potential of connected vehicle technologies to enable proactive rerouting under roadway incidents and congestion-prone conditions, thereby reducing delays and improving safety outcomes \cite{Ye2022,11312449,ho2019improved}. Concurrently, eco-routing and multi-objective optimization strategies have shown that intelligent route selection can reduce energy consumption and environmental impact in connected vehicle environments \cite{Alfaseeh2020}. Dynamic multi-objective eco-routing studies have further demonstrated the value of incorporating real-time traffic and emission information into CAV routing decisions \cite{djavadian2020multiobjective}. Recent cooperative CAV routing and traffic-organization methods also show that distributed, intersection-level, and network-level decisions can improve traffic dispersion, safety, and energy efficiency in urban networks \cite{vitale2024cooperative,wang2020cooperativeOrganization}. Moreover, large-scale dynamic rerouting simulations have illustrated system-level performance impacts when predictive routing strategies are deployed at metropolitan scales \cite{Chan2023}. While these studies demonstrate the value of adaptive routing, they generally rely on reactive or partially predictive information and do not fully exploit modern data-driven forecasting capabilities.

Recent advances in machine learning, particularly graph neural networks for traffic forecasting, have significantly improved the ability to model and predict traffic dynamics in complex urban networks \cite{jiang2022survey}. Spatiotemporal deep learning models, in particular, have demonstrated strong performance in capturing both the spatial dependencies between road segments and the temporal evolution of traffic conditions. Graph-based forecasting models such as spatio-temporal graph convolutional networks (STGCN) demonstrate that road-network traffic can be modeled as graph-structured time series to capture spatial and temporal dependencies \cite{yu2018stgcn}. Similarly, adaptive graph-based models such as Graph WaveNet and adaptive graph convolutional recurrent networks show that hidden spatial dependencies can be learned directly from traffic data rather than relying only on fixed road-network connectivity \cite{wu2019graphwavenet,bai2020adaptive}. Among these approaches, diffusion convolutional recurrent neural networks (DCRNN) and attention-based graph forecasting models provide powerful frameworks for modeling traffic flow on graph-structured networks, enabling accurate multi-step forecasting of congestion patterns \cite{li2018dcrnn,zheng2020gman}. Extensions and reproducibility studies of these models further highlight their effectiveness and practical applicability in real-world traffic systems \cite{sun2023diffusion}. By learning how congestion propagates across connected road segments over time, including dynamic long-range effects, propagation-delay effects, and local fluctuations caused by traffic accidents, such models offer the potential to anticipate disruptions before they occur \cite{jiang2023pdformer,ye2023dynamic,kinchen2026spatiotemporal}. This predictive capability is essential for transitioning from reactive traffic management strategies to proactive control mechanisms that can influence vehicle behavior in advance of congestion formation.

In parallel, anomaly, incident, and crash prediction methods have been developed to identify irregular traffic patterns and collision-related disruptions using statistical, deep learning, and graph-based approaches \cite{zhao2021prediction,li2021stgcn,yu2021deep}. Spatial-temporal graph and hypergraph neural network methods have shown that traffic anomalies can be detected by comparing predicted traffic states with observed traffic states, modeling sparse anomalies, and capturing their spreading effects across neighboring road segments \cite{zhang2022automatic,feng2024traffic}. Recent crash prediction research has also demonstrated that probabilistic graph neural networks can estimate road-level crash risk while accounting for uncertainty in spatially and temporally sparse crash data \cite{gao2024uncertainty}. Comprehensive surveys of traffic incident detection methods further highlight the diversity of approaches and the importance of accurate detection for mitigating non-recurrent congestion \cite{elsahly2022systematic,chew2024survey}. These methods aim to detect deviations from normal traffic behavior, such as sudden slowdowns, abnormal flow patterns, or incident-induced disturbances. While such techniques improve situational awareness and can provide early warnings of potential disruptions, they are typically designed as standalone detection systems. As a result, their outputs are not directly integrated into routing or control decisions, limiting their practical impact on real-time traffic operations. Bridging this gap between prediction, detection, and control remains a critical challenge in intelligent transportation systems.

Despite these advances, existing approaches often focus on either congestion mitigation, safety improvement, or emissions reduction in isolation. Similarly, predictive models for traffic forecasting and anomaly detection are rarely integrated directly into real-time traffic control frameworks. Few studies consider how predictive artificial intelligence (AI) models can be embedded within microscopic traffic simulators to proactively reroute CAVs in response to anticipated congestion and collision events while simultaneously accounting for environmental impact. This lack of integration limits the ability of current systems to fully leverage predictive intelligence for coordinated, system-level traffic management. Addressing this gap requires a unified framework that combines predictive modeling, simulation-based evaluation, and real-time routing control in a cohesive and scalable manner.

The main contributions of this paper are summarized as follows:
\begin{itemize}
    \item A collision-aware predictive rerouting framework is developed for connected and automated vehicles operating under non-recurrent congestion and incident-induced traffic disruptions.

    \item A DCRNN-based spatiotemporal traffic forecasting model is integrated with microscopic SUMO simulation through TraCI to estimate future congestion conditions on graph-structured road networks.

    \item A multi-objective routing strategy is formulated to jointly account for predicted congestion, $\mathrm{CO}_2$ emissions, route length, and collision-window avoidance in real-time CAV rerouting decisions.

    \item The proposed framework is evaluated against benchmark rerouting strategies, including K-shortest path routing, predictive density-based routing, V2X proactive routing, SUMO-based rerouting, and no-rerouting baselines across multiple CAV penetration levels.
\end{itemize}

The remainder of this paper is organized as follows. Section~II presents the problem formulation, including the network representation, traffic-state model, collision-aware constraints, and decentralized routing objective. Section~III describes the proposed simulation framework, predictive traffic-state modeling approach, collision-aware rerouting mechanism, benchmark methods, and performance evaluation procedure. Section~IV presents and discusses the simulation results. Finally, Section~V concludes the paper and outlines directions for future work.

\section{Problem Formulation}\label{Section II}

This section states the routing problem considered in this work. The implementation details, training configuration, simulation settings, and evaluation procedure are provided in Section~\ref{Section III}.

\subsection{Network and Traffic-State Representation}

The transportation network is represented as a directed graph $G=(V,A)$, where $V$ is the set of normal road segments and $A$ is the set of feasible directed transitions between segments. Each road segment $e\in V$ has length $L(e)>0$ and time-varying traffic attributes obtained from the microscopic simulation.

Simulation time is discrete. Let $t=n\Delta t$, where $n\in\mathbb{Z}_{\geq 0}$ and $\Delta t$ is the simulation step length. At each time step, the edge-level traffic state is represented by a feature matrix $X_t\in\mathbb{R}^{|V|\times F}$. In the implemented framework, each row corresponds to one road segment and contains the configured traffic features used for online inference, including the live Travel Time Index (TTI), $\mathrm{CO}_2$ emissions, average speed, vehicle count, lane occupancy, collision-window indicators, and spike indicators. These quantities are distinguished from the evaluation metrics, which are computed after simulation from matched vehicle trajectories.

\subsection{Prediction-Assisted Routing Objective}

For each eligible CAV $k$, the routing decision is the selected route $\mathcal{R}_k\in\mathcal{P}_k$, where $\mathcal{P}_k$ denotes the feasible route set from the current edge to the destination or to an admissible fallback endpoint. The objective is to select a feasible route that reduces predicted congestion, limits emissions, avoids excessive detours, and avoids active or imminent incident-affected regions. In compact form, the routing objective is written as
\begin{equation*}
    \mathcal{R}_k^*
    \in
    \arg\min_{\mathcal{R}\in\mathcal{P}_k}
    C(\mathcal{R};t),
\end{equation*}
subject to route feasibility, collision-window avoidance, detour limits, emission-improvement gates, cooldown constraints, and reroute-count limits.

\subsection{Incident-Aware Feasibility}

Let $\mathcal{A}(t)\subseteq V$ denote the set of road segments affected by active incident windows at time $t$. Because the controller also considers imminent disruptions, the implementation uses a finite lookahead set $\mathcal{L}_k(t)$ for each CAV, defined from the remaining route and the predicted arrival times of the vehicle on upcoming edges. A candidate route is considered incident-feasible when its downstream edges do not overlap the active or lookahead incident sets within the relevant decision horizon. This formulation avoids treating incident information as an additive cost only; instead, it rejects routes that would continue through active collision-window regions when avoidance is triggered.

\subsection{Evaluation Objective}

The evaluation compares routing methods under matched simulation demand, origin--destination structure, incident schedules, CAV penetration levels, and random seeds where specified. Trip-level outcomes are the primary quantities of interest, including total travel time, total distance, total $\mathrm{CO}_2$ emissions, completion status, and incident exposure. Edge-level TTI is retained as a diagnostic metric for congestion exposure but is not used as the sole basis for route-level performance claims.

\section{Methodology}\label{Section III}

\subsection{Simulation Framework and Scenario Construction}

The proposed framework integrates microscopic traffic simulation, live traffic-state estimation, DCRNN-based congestion forecasting, collision-window awareness, and adaptive rerouting of CAVs within a unified control architecture. The system is implemented using the Simulation of Urban Mobility (SUMO) platform and controlled through the TraCI interface, enabling real-time interaction with the simulation environment.

The transportation network is modeled as a directed edge-transition graph. In this representation, each graph node corresponds to a normal SUMO road segment, and each directed graph connection corresponds to a feasible transition from one road segment to another. Internal junction edges are excluded from the normal routing graph where appropriate, although they may still be used internally by SUMO for vehicle movement and collision-stop placement.

The overall workflow consists of four primary stages. First, route generation and scenario construction are performed to define traffic demand, vehicle types, and programmable collision events. Second, no-collision baseline route files are generated by removing collision-forcing stop commands while preserving the original vehicle demand and route structure. Third, single-vehicle no-collision baseline simulations are executed to obtain reference edge-level metrics. Finally, collision-enabled simulations are conducted under no-rerouting, benchmark-rerouting, and DCRNN-based predictive-rerouting configurations. A post-processing stage evaluates system performance by comparing matched vehicle trajectories across scenarios.

At each simulation step, TraCI is used to extract vehicle-level and edge-level metrics, including vehicle position, speed, route index, edge vehicle count, lane occupancy, and instantaneous $\mathrm{CO}_2$ emissions. These data serve as inputs to the predictive traffic-state model and are also logged for performance evaluation. Traffic demand is generated using a graph-based route construction process. The road network topology is extracted from SUMO network files and represented as a directed adjacency structure. Valid entry and exit edges are identified, and feasible routes are constructed subject to connectivity and path-length constraints. Vehicle populations include passenger vehicles, buses, and connected and automated vehicles. In the implemented CAV penetration experiments, penetration is produced by converting eligible non-collision passenger vehicles into CAVs until the desired penetration level is reached. Thus, CAV penetration changes the composition of the vehicle fleet while preserving the overall traffic demand and route structure.

Programmable collision scenarios are embedded directly into the generated route definitions. Two primary collision types are considered: intersection collisions and rear-end collisions. Intersection collisions are constructed by coordinating two vehicles near a shared junction conflict region and forcing one or both vehicles to stop in or near the junction area. Rear-end collisions are generated near bus stops or constrained segments, where a bus stop and following vehicles create queue-forming disruption conditions.

Each collision event is associated with metadata specifying the affected edge or edges, activation time, stop duration, involved vehicle identifiers, and collision type. This information is stored externally and used during simulation to define dynamic collision windows. The use of programmable and repeatable collision scenarios enables systematic evaluation of rerouting strategies under controlled non-recurrent congestion conditions.

Table~\ref{tab:scenario_summary} summarizes the simulation networks and output logs used in the current study. The experiment uses the NYC--6 Intersections, Houston Street Playground, and Penn South SUMO networks. The available exported comparison workbooks contain 15{,}535--21{,}424 vehicle-edge records, 529--777 unique vehicle identifiers, and 60--92 unique traversed road-segment identifiers per scenario output. The collision-event logs contain two rear-end-bus scenarios with 80 bus-stop disruption records each and one mixed rear-end scenario with 80 bus-stop disruption records and 54 rear-end vehicle-pair events. Across the available logs, incident start times span from 343~s to 24{,}929~s, and the maximum observed simulation timestamp in the exported workbooks is 25{,}837~s. CAV penetration levels of 5\%, 15\%, 25\%, 35\%, and 50\% are evaluated by converting eligible non-collision passenger vehicles into CAVs while preserving the underlying route demand.

\begin{table*}[t]
\centering
\caption{Available Simulation Scenario and Output-Log Summary}
\label{tab:scenario_summary}
\small
\begingroup
\setlength{\tabcolsep}{2.5pt}
\renewcommand{\arraystretch}{1.05}
\resizebox{\textwidth}{!}{%
\begin{tabular}{@{}lcccccc@{}}
\toprule
Item & Scenario output 1 & Scenario output 2 & Scenario output 3 & Combined / policy & Source used in evaluation & Notes \\
\midrule
Network families & \multicolumn{3}{c}{NYC--6 Intersections, Houston Street Playground, and Penn South} & three maps & SUMO network files & map-specific XML topology files \\
Vehicle-edge records & 21{,}424 & 17{,}614 & 15{,}535 & 54{,}573 total & comparison workbooks & records after header row \\
Unique vehicle identifiers & 529 & 775 & 777 & 529--777 per output & comparison workbooks & includes CAV, passenger, and bus vehicles \\
Unique traversed edge identifiers & 60 & 60 & 92 & 60--92 per output & comparison workbooks & observed road-segment IDs in logged trajectories \\
Incident records & 80 & 80 & 188 & 348 rows & collision logs & rear-end-bus and rear-end records \\
Grouped incident events & 80 & 80 & 134 & 294 events & collision logs & rear-end pairs grouped by common start time \\
Incident start-time range & 523--24{,}082~s & 343--24{,}019~s & 1{,}036--24{,}929~s & 343--24{,}929~s & collision logs & programmed disruption times \\
Incident edge identifiers & 2 & 2 & 1 & 5 listed IDs & collision logs & collision-window affected edges \\
CAV selection rule & \multicolumn{3}{c}{eligible non-collision passenger vehicles converted to CAVs} & fixed demand & route-generation process & preserves base route demand \\
Penetration levels & \multicolumn{3}{c}{5\%, 15\%, 25\%, 35\%, and 50\%} & five levels & penetration route files & CAV-only and aggregate plots reported \\
Simulation step length & \multicolumn{3}{c}{1~s} & fixed & runtime SUMO configuration & begin time set to 0~s \\
Teleport handling & \multicolumn{3}{c}{disabled using \texttt{--time-to-teleport -1}} & fixed & runtime SUMO command & disconnected teleport also disabled \\
\bottomrule
\end{tabular}%
}
\endgroup
\end{table*}

For reproducibility, the scenario table reports the quantities used in the present evaluation from the exported workbooks, collision logs, and runtime scripts. The runtime SUMO network files provide the topology used by the simulations, while the exported trajectory logs provide the observed vehicle-edge records and traversed-edge counts reported in Table~\ref{tab:scenario_summary}.

\subsection{Baseline Evaluation and Predictive Traffic-State Modeling}

To quantify the impact of congestion and rerouting, a no-collision baseline is constructed from the collision-enabled route files. Collision-forcing stops are removed from intersection collision vehicles and from the lead and rear vehicles in rear-end collision scenarios. However, bus-stop dwell behavior for bus collision vehicles is retained so that normal bus-stop operations remain represented in the baseline.

Each vehicle is then simulated independently in SUMO to obtain baseline edge-level metrics without congestion or collision interference. During these single-vehicle baseline simulations, the temporary vehicle departure time is normalized to $t=0$ so that baseline measurements capture route traversal performance rather than idle simulation time before the original scheduled departure.

For vehicle $v$ traversing edge $e$, the evaluation Travel Time Index (TTI) is computed using observed simulation step counts:
\begin{equation}
    \mathrm{TTI}^{\mathrm{eval}}_{v,e}
    =
    \frac{
        S^{\mathrm{with}}_{v,e}
    }{
        S^{\mathrm{base}}_{v,e}
    },
    \label{eq:tti_eval}
\end{equation}
where $S^{\mathrm{with}}_{v,e}$ denotes the number of simulation steps spent on edge $e$ during the collision-enabled simulation, and $S^{\mathrm{base}}_{v,e}$ denotes the corresponding no-collision baseline value.

Since the simulation step length is one second in the implemented runtime configuration, the step-count ratio in \eqref{eq:tti_eval} is equivalent to an edge-level travel-time ratio. Values greater than one indicate degraded travel conditions relative to the no-collision baseline.

For live routing and DCRNN feature construction, a separate speed-based TTI feature is also used:
\begin{equation}
    \mathrm{TTI}^{\mathrm{live}}(e,t)
    =
    \frac{
        v_{\mathrm{free}}(e)
    }{
        \max(0.1,v(e,t))
    },
    \label{eq:tti_live}
\end{equation}
where $v_{\mathrm{free}}(e)$ is the free-flow speed for edge $e$ and $v(e,t)$ is the observed mean speed on edge $e$ at simulation time $t$. The lower bound in the denominator is a numerical safeguard against division by very small measured speeds during stops and does not redefine free-flow TTI.

Future traffic conditions are estimated using a DCRNN, which captures spatial dependencies in the transportation network and temporal evolution of traffic states. The DCRNN uses the directed edge-transition graph defined in Section~\ref{Section II}; in this graph, each node corresponds to a SUMO road segment and each directed arc represents a feasible transition between road segments.

At each simulation time step $t$, an edge-level feature matrix
\begin{equation*}
    X_t \in \mathbb{R}^{|V| \times F}
\end{equation*}
is constructed using real-time data obtained from TraCI. The implemented feature vector is
\begin{equation*}
    x_e(t)=
    \big[
    \mathrm{TTI}^{\mathrm{live}},
    \mathrm{CO}_2,
    v,
    n,
    o,
    i_c,
    i_s
    \big]^{\top},
\end{equation*}
where $v$ is average speed, $n$ is vehicle count, $o$ is occupancy, $i_c$ is the collision-window indicator, and $i_s$ is the spike indicator. The corresponding units are dimensionless for TTI and indicators, milligrams per second for instantaneous $\mathrm{CO}_2$, meters per second for speed, vehicles for count, and percent for occupancy.

The live feature matrix is normalized using stored scaler parameters:
\begin{equation}
    \widetilde{X}_t
    =
    (X_t-\mu)\oslash(\sigma+\varepsilon_s),
    \label{eq:feature_normalization}
\end{equation}
where $\mu$ and $\sigma$ are feature-wise training means and scales, $\oslash$ denotes element-wise division with feature-wise broadcasting, and $\varepsilon_s$ is a small numerical constant.

A rolling lookback sequence is maintained:
\begin{equation*}
    \mathcal{X}_t
    =
    \left[
        \widetilde{X}_{t-\ell+1},
        \ldots,
        \widetilde{X}_t
    \right],
\end{equation*}
where $\ell$ is the DCRNN lookback length. The symbol $\mathcal{X}_t$ is used for the input sequence to avoid confusion with the recurrent hidden state.

Let $W$ denote the DCRNN adjacency matrix. The implementation uses a row-normalized directed transition matrix
\begin{equation}
    \overline{W}_{ij}
    =
    \frac{W_{ij}}
         {\sum_j W_{ij}+10^{-6}}.
    \label{eq:row_normalized_adjacency}
\end{equation}
This one-direction diffusion operator is used consistently in the implemented controller to propagate information over feasible edge transitions.

The diffusion convolution operation can be written as
\begin{equation}
    \mathrm{DiffConv}(Z)
    =
    \sum_{k=0}^{K-1}
    \overline{W}^{k}Z\Theta_k,
    \label{eq:diffusion_conv}
\end{equation}
where $K$ is the diffusion order and $\Theta_k$ are learned parameters.

The recurrent update is expressed as
\begin{align*}
    Z_t &= [\widetilde{X}_t,H_{t-1}], \\
    U_t &= \mathrm{DiffConv}(Z_t), \\
    H_t &= \mathrm{GRU}(U_t,H_{t-1}),
\end{align*}
where $H_t$ denotes the recurrent hidden state.

For a forecast horizon of $H_f$ steps, the model output is represented as
\begin{equation}
    (Y_{t+1:t+H_f}^{-},Y_{t+1:t+H_f}^{+})
    =
    \mathcal{F}_{\mathrm{DCRNN}}(\mathcal{X}_t,G),
    \label{eq:dcrnn_horizon_output}
\end{equation}
where $Y^{-}$ and $Y^{+}$ denote the lower and upper output tensors, respectively. In the routing implementation, the midpoint is used as the forecast supplied to the route-cost function:
\begin{equation*}
    \widehat{Y}_{t+1:t+H_f}
    =
    \frac{Y_{t+1:t+H_f}^{-}+Y_{t+1:t+H_f}^{+}}{2}.
\end{equation*}
In the implemented controller, the lower and upper outputs are retained for architectural compatibility, and the midpoint is used as the routing forecast. Interval calibration, empirical coverage, and interval-width analysis are reserved for future uncertainty-aware extensions.

The first inverse-scaled output channel is used as the predicted TTI:
\begin{equation}
\begin{aligned}
    \widehat{\mathrm{TTI}}(e,t+h)
    =
    \mathrm{clip}\big(
        &\sigma_{\mathrm{TTI}}\widehat{Y}_{e,h,1}
        +
        \mu_{\mathrm{TTI}}, \\
        &0.05,
        20
    \big),
\end{aligned}
\label{eq:dcrnn_tti_prediction}
\end{equation}
where $h\in\{1,\ldots,H_f\}$ is the forecast step. The clipping interval is used as a numerical guard for model outputs; evaluation TTI remains defined by \eqref{eq:tti_eval}.

The DCRNN prediction therefore provides the congestion forecast used in route scoring. The emission term used in routing is computed from live or fallback edge-emission values rather than directly from a DCRNN-predicted $\mathrm{CO}_2$ output.

\begin{table*}[t]
\centering
\caption{Runtime DCRNN Inference Configuration Used by the Routing Controller}
\label{tab:dcrnn_config}
\small
\begingroup
\renewcommand{\arraystretch}{1.12}
\begin{tabular}{@{}p{0.27\textwidth}p{0.68\textwidth}@{}}
\toprule
Item & Runtime value or implementation detail \\
\midrule
Model artifacts loaded online & \texttt{trained\_dcrnn.pt}, \texttt{W\_filtered.npy}, \texttt{edges\_filtered.json}, and \texttt{scaler\_params.json} \\
Input feature order & read from \texttt{scaler\_params.json}; common pipeline order is TTI, \texttt{total\_co2\_mg\_with}, \texttt{avg\_speed\_mps\_with}, \texttt{collision\_flag}, \texttt{tti\_spike\_flag}, and \texttt{co2\_spike\_flag} \\
Online live measurements & edge mean speed, vehicle count, lane occupancy, edge $\mathrm{CO}_2$ emission, and collision-stop flag from TraCI \\
Feature scaling & feature-wise mean and scale from \texttt{scaler\_params.json}; $\widetilde{X}_t=(X_t-\mu)\oslash(\sigma+\varepsilon_s)$ \\
Lookback length $\ell$ & read from \texttt{dcrnn\_training\_settings.json}; runtime fallback is 5 live simulation steps \\
Forecast output used online & one live DCRNN output tensor per edge; the midpoint of lower and upper outputs is used for routing \\
Hidden dimension & read from \texttt{dcrnn\_training\_settings.json}; runtime fallback is 64 \\
Number of recurrent layers & one DCRNN recurrent cell with a GRUCell update in the runtime implementation \\
Diffusion order $K$ & read from \texttt{dcrnn\_training\_settings.json}; runtime fallback is 2 \\
Diffusion operator & powers of one row-normalized adjacency matrix $\overline{W}_{ij}=W_{ij}/(\sum_j W_{ij}+10^{-6})$ \\
Output construction & two linear heads produce lower and upper outputs; a softplus half-width is applied and the midpoint is routed on \\
TTI clipping for routing & predicted TTI is clipped to $[0.05,20]$ as a numerical guard; evaluation TTI is computed separately from baseline traversal steps \\
Offline training metadata & not required during online inference; the controller loads the stored model weights, graph, edge list, and scaler parameters \\
\bottomrule
\end{tabular}
\endgroup
\end{table*}

\subsection{Collision-Aware Multi-Objective Rerouting}

To explicitly represent non-recurrent congestion caused by incidents, the framework defines time-dependent collision windows for affected edges. Each programmed collision event $c$ is associated with a start time, end time, collision type, and set of impacted edges.

Let $\mathcal{E}_c$ denote the set of edges affected by collision event $c$. The active collision-edge set at time $t$ is
\begin{equation}
    \mathcal{A}(t)
    =
    \bigcup_{c:\,t\in W_c}
    \mathcal{E}_c,
    \label{eq:active_collision_edges}
\end{equation}
where $W_c$ is the active time window for event $c$.

A finite lookahead mechanism is employed to anticipate future conflicts before CAVs reach affected regions. During each simulation step, the routing controller examines the remaining route of each CAV and determines whether any upcoming edges are part of an active or imminent collision window. In addition to $\mathcal{A}(t)$, the controller forms a lookahead set $\mathcal{L}_k(t)$ containing downstream edges whose estimated arrival times overlap a scheduled collision window for vehicle $k$.

If a potential conflict is detected, the vehicle is flagged for rerouting consideration at the next feasible decision point. Candidate reroutes that still contain active collision-window edges ahead of the current edge are rejected. Thus, collision awareness is enforced through active-window lookahead and route blocking rather than through an additive collision-severity cost term.

Routing decisions are based on a multi-objective cost function that captures predicted congestion, environmental impact, and route length. For a candidate route $\mathcal{R}$, the implemented route cost is written as
\begin{equation}
\begin{aligned}
    C(\mathcal{R};t)
    =
    \sum_{e \in \mathcal{R}}
    \big(&W_{\mathrm{TTI}}(e,t)
        +W_{\mathrm{CO}_2}(e,t)\\
        &+W_L(e)\big),
\end{aligned}
\label{eq:route_cost_weighted_terms}
\end{equation}
where
\begin{align}
    W_{\mathrm{TTI}}(e,t)
    &= \alpha\,\widehat{\mathrm{TTI}}(e,t), \\
    W_{\mathrm{CO}_2}(e,t)
    &= \gamma\,\widetilde{\mathrm{CO}_2}(e,t), \\
    W_L(e)
    &= \delta\,\widetilde{L}(e).
\end{align}

Here, $\widehat{\mathrm{TTI}}(e,t)$ is the DCRNN-predicted congestion term, $\widetilde{\mathrm{CO}_2}(e,t)$ is the normalized live or proxy edge-emission term, and $\widetilde{L}(e)$ is the normalized edge-length term. The DCRNN implementation uses
\begin{equation*}
    \alpha=0.95,
    \qquad
    \gamma=3.1,
    \qquad
    \delta=0.90.
\end{equation*}
The weighted sum in \eqref{eq:route_cost_weighted_terms} is evaluated over downstream candidate edges. Because edge segmentation can influence a pure edge-sum objective, route-level reporting also includes total travel time, distance, and emissions.

The normalized emission term is computed as
\begin{equation}
    \widetilde{\mathrm{CO}_2}(e,t)
    =
    \frac{
        \mathrm{CO}_2(e,t)-\mathrm{CO}_{2,\min}(t)
    }{
        \mathrm{CO}_{2,\max}(t)
        -
        \mathrm{CO}_{2,\min}(t)
        +
        \varepsilon_{\mathrm{CO}_2}
    }.
    \label{eq:co2_normalization}
\end{equation}
and the normalized length term is computed as
\begin{equation}
    \widetilde{L}(e)
    =
    \frac{
        L(e)-L_{\min}
    }{
        L_{\max}-L_{\min}+\varepsilon_L
    }.
    \label{eq:length_normalization}
\end{equation}

The route-search implementation initializes the current edge with zero search cost and adds the weighted cost of downstream candidate edges as the graph search expands. For reporting and route comparison, the route-metric computation evaluates the weighted terms over the route passed to the metric function. Therefore, \eqref{eq:route_cost_weighted_terms} represents the route-level scoring structure, while the graph-search implementation applies the same edge-weight terms during downstream path expansion.

Rerouting decisions are applied only to CAVs and are evaluated at discrete time intervals. In the DCRNN penetration experiment, all existing CAVs in the penetration route files are controlled by the DCRNN-based routing method.

The framework employs an intersection-aware strategy in which rerouting is primarily evaluated near feasible decision points. For each candidate reroute, the controller compares the predicted cost of the current route with the cost of an alternative route. The relative improvement is computed as
\begin{equation}
    \mathrm{Imp}
    =
    \frac{
        C_{\mathrm{cur}} - C_{\mathrm{alt}}
    }{
        \max(C_{\mathrm{cur}}, \varepsilon)
    }.
    \label{eq:relative_improvement}
\end{equation}

A reroute is applied only if the alternative route satisfies additional acceptance conditions. These include route validity, collision-window safety, cooldown periods, reroute-count limits, maximum allowable route-length increase, and emission-improvement constraints. In the implemented DCRNN configuration, normal reroutes are constrained by
\begin{equation}
    L_{\mathrm{alt}}
    \le
    1.12\,L_{\mathrm{cur}},
    \label{eq:length_gate}
\end{equation}
and by an emission-improvement condition of the form
\begin{equation}
\begin{aligned}
    \mathrm{CO}_{2,\mathrm{alt}}
    &\le
    (1-\eta_{\mathrm{CO}_2})
    \mathrm{CO}_{2,\mathrm{cur}}, \\
    \eta_{\mathrm{CO}_2}
    &=0.003,
\end{aligned}
\label{eq:co2_gate}
\end{equation}
except in explicitly configured target-fill fallback cases.

Near intersections, candidate outgoing branches may also be scored using predicted TTI, live vehicle count, occupancy, normalized $\mathrm{CO}_2$, and an empty-road bonus. This intersection-aware branch scoring helps align rerouting decisions with physically feasible network choices.

\begin{table}[t]
\centering
\caption{Routing Gate and Hyperparameter Summary for the Proposed Method}
\label{tab:routing_gates}
\footnotesize
\resizebox{\columnwidth}{!}{%
\begin{tabular}{@{}ll@{}}
\toprule
Parameter & Value / selection rule \\
\midrule
TTI weight $\alpha$ & 0.95 \\
Emission weight $\gamma$ & 3.1 \\
Length weight $\delta$ & 0.90 \\
TTI rerouting threshold & 1.03 \\
Rerouting update interval & 1.0~s \\
Minimum relative improvement & 0.00 \\
Minimum absolute route-cost drop & 0.00 \\
Minimum total TTI drop & 0.00 \\
Minimum maximum-edge TTI drop & 0.00 \\
Normal maximum detour factor & 1.12 \\
Emission-improvement threshold $\eta_{\mathrm{CO}_2}$ & 0.003 \\
Periodic CO$_2$ acceptance margin & 0.006 \\
Cooldown period & 4.0~s \\
Maximum reroutes per CAV & 3 \\
Non-forced reroute fraction & 1.00 \\
Maximum non-forced reroutes per interval & 10 \\
Target reroutes per map & 1--3 \\
Target-fill maximum CO$_2$ increase & 1.04$\times$ current-route estimate \\
Target-fill maximum length ratio & 1.20$\times$ current-route length \\
Collision path lookahead & 4 downstream edges \\
Collision blocking horizon & 4 downstream edges \\
Forced active-collision horizon & 8 downstream edges \\
Collision lookahead time & 180~s \\
Intersection route-cost drop & 0.00 \\
Intersection next-edge TTI drop & 0.00 \\
Intersection branch gap & 0.00 \\
Empty-road bonus & 0.01 \\
Vehicle-count weight at intersections & 0.10 \\
Occupancy weight at intersections & 0.08 \\
Stress persistence & 1 simulation step \\
\bottomrule
\end{tabular}%
}
\end{table}

\subsection{Benchmark Methods and Performance Evaluation}

To evaluate the effectiveness of the proposed DCRNN rerouting framework, multiple benchmark rerouting strategies are implemented within the same SUMO-based simulation environment. The information and constraints available to each method are summarized in Table~\ref{tab:method_inputs}. The comparison uses matched demand, incident schedules, penetration levels, and random seeds, while documenting method-specific information and acceptance constraints in Table~\ref{tab:method_inputs}.

\begin{table*}[t]
\centering
\caption{Routing Information and Constraints Available to Each Method}
\label{tab:method_inputs}
\small
\begingroup
\setlength{\tabcolsep}{3pt}
\renewcommand{\arraystretch}{1.18}
\begin{tabular}{@{}L{0.15\textwidth}L{0.24\textwidth}L{0.27\textwidth}L{0.25\textwidth}@{}}
\toprule
Method & Information available & Candidate-route behavior & Acceptance gates and route weights \\
\midrule
Proposed DCRNN method & Learned forecast: Yes; live traffic state: Yes; incident-window information: active and lookahead windows & DCRNN weighted graph search to destination or admissible fallback & Acceptance gates: Table~\ref{tab:routing_gates}; route weights: $(0.95,3.1,0.90)$ \\
K-shortest path & Learned forecast: No; live traffic state: Yes; incident-window information: 8-edge / 30-s collision lookahead & limited K-shortest feasible candidates & Acceptance gates: 20\% relative, 1.50 absolute, 60-s cooldown, 6 reroutes; route weights: $(1.2,1.0,0.6)$ \\
Predictive density & Learned forecast: No; live traffic state: Yes; incident-window information: 8-edge / 30-s collision lookahead & live-cost routing with density prediction penalty & Acceptance gates: 20\% relative, 1.50 absolute, 60-s cooldown, 6 reroutes; route weights: $(1.2,1.0,0.6)$ plus $k_c=80$~veh/km \\
V2X proactive & Learned forecast: No; live traffic state: Yes; incident-window information: 180-s broadcast and 8-edge lookahead & incident-aware live-cost routing & Acceptance gates: 20\% relative, 1.50 absolute, 60-s cooldown, 6 reroutes; route weights: $(1.2,1.0,0.6)$ \\
SUMO-based reroute & Learned forecast: No learned DCRNN controller; live traffic state: Yes; incident-window information: valid-exit routing reference & SUMO/valid-exit route selection at 10-s intervals & Acceptance gates: stop when $\leq 2$ edges remain; route weights: SUMO route selection \\
No reroute & Learned forecast: No; live traffic state: logged only; incident-window information: No rerouting response & original programmed route & Acceptance gates: none; route weights: none \\
\bottomrule
\end{tabular}
\endgroup
\end{table*}

\subsubsection{Basic SUMO Rerouting}

The Basic SUMO benchmark uses SUMO-based rerouting behavior without the DCRNN prediction model. This method relies on current network conditions and feasible exit routing rather than learned spatiotemporal forecasting. It provides a baseline for comparing the benefits of the proposed predictive DCRNN controller.

\subsubsection{K-Shortest Path Rerouting}

The K-shortest path benchmark generates a limited set of feasible candidate paths and selects the path with the lowest live route cost. The implemented benchmark cost follows the same general weighted-term structure:
\begin{equation}
\begin{aligned}
    C_{\mathrm{KSP}}(\mathcal{R};t)
    =
    \sum_{e \in \mathcal{R}}
    \big(
        &W^{b}_{\mathrm{TTI}}(e,t)
        +
        W^{b}_{\mathrm{CO}_2}(e,t) \\
        &+
        W^{b}_{L}(e)
    \big),
\end{aligned}
\label{eq:ksp_cost}
\end{equation}
where
\begin{align}
    W^{b}_{\mathrm{TTI}}(e,t)
    &= \alpha_b\,\mathrm{TTI}^{\mathrm{live}}(e,t), \\
    W^{b}_{\mathrm{CO}_2}(e,t)
    &= \gamma_b\,\widetilde{\mathrm{CO}_2}(e,t), \\
    W^{b}_{L}(e)
    &= \delta_b\,\widetilde{L}(e).
\end{align}

The benchmark weights are
\begin{equation*}
    \alpha_b=1.2,
    \qquad
    \gamma_b=1.0,
    \qquad
    \delta_b=0.6.
\end{equation*}
Unlike the proposed DCRNN framework, this benchmark does not use learned DCRNN forecasts.

\subsubsection{Predictive Density-Based Rerouting}

The predictive density benchmark estimates near-future congestion by analyzing temporal changes in edge-level density. For edge $e$, density is computed as
\begin{equation}
    k(e,t)
    =
    \frac{
        n(e,t)
    }{
        L(e)/1000
    },
    \label{eq:density}
\end{equation}
where $n(e,t)$ is the number of vehicles on edge $e$ and $L(e)$ is measured in meters. Thus, $k(e,t)$ is expressed in vehicles per kilometer.

The density growth rate is computed as
\begin{equation}
    r(e,t)
    =
    \frac{
        k(e,t)-k(e,t-\Delta t)
    }{
        \Delta t
    }.
    \label{eq:density_growth}
\end{equation}

The estimated time to reach critical density is
\begin{equation}
t_c(e,t)
=
\begin{cases}
0,
& k(e,t)\ge k_c,\\[4pt]
\infty,
& \substack{
    r(e,t)\le 0\\
    k(e,t)<k_c
  },\\[6pt]
\dfrac{k_c-k(e,t)}
      {\max(r(e,t),10^{-6})},
& \text{otherwise}.
\end{cases}
\label{eq:time_to_critical}
\end{equation}

In the implemented benchmark, the critical density is
\begin{equation*}
    k_c=80\,\mathrm{veh/km}.
\end{equation*}
Edges predicted to approach critical density within the configured threshold window are assigned additional routing penalties. This approach provides short-term anticipation of congestion formation but does not explicitly model graph-based spatiotemporal dependencies using a learned DCRNN.

\subsubsection{V2X Proactive Rerouting}

The V2X proactive benchmark uses direct incident-awareness information from scheduled collision windows. Vehicles receive information regarding active or imminent collision regions through simulated connected-vehicle communication.

When an upcoming route segment is identified as part of an active or imminent collision region, vehicles proactively reroute to avoid the affected area. In the implemented benchmark, V2X routing also considers live traffic conditions and route-cost terms involving TTI, normalized $\mathrm{CO}_2$, and normalized length. However, it does not use DCRNN-predicted congestion forecasts.

\subsubsection{Comparison Framework}

All methods are evaluated using matched simulation inputs, including collision schedules, network topology, vehicle demand, and CAV penetration levels, subject to the information availability summarized in Table~\ref{tab:method_inputs}. Performance is assessed using matched vehicle trajectories and route-level metrics, including total simulation time, distance traveled, average speed, congestion exposure, completion status, and total $\mathrm{CO}_2$ emissions.

The proposed DCRNN framework is compared against these benchmark methods to assess the benefits of integrating spatiotemporal prediction, collision-window lookahead mechanisms, and multi-objective route scoring within a unified rerouting architecture. The simulation campaign also includes ablation variants with live-state routing only, collision-window routing without DCRNN forecasting, DCRNN routing without collision-window information, and the full DCRNN collision-window framework.

Comprehensive data logging is performed during simulation to capture both vehicle-level and system-level performance. Metrics include simulation step count, distance traveled, average speed, and total $\mathrm{CO}_2$ emissions for each matched vehicle record, as well as congestion and collision-window indicators.

The primary evaluation uses a trip-level matched design. For each vehicle, the origin, destination, departure time, incident scenario, CAV identity, and random seed are held fixed across methods whenever possible. For each matched vehicle $v$, changes in trip-level performance metrics are computed as
\begin{align}
    \Delta S_{v}
    &=
    S^{\mathrm{reroute}}_{v}
    -
    S^{\mathrm{no\mbox{-}reroute}}_{v},
    \label{eq:delta_steps}\\
    \Delta D_{v}
    &=
    D^{\mathrm{reroute}}_{v}
    -
    D^{\mathrm{no\mbox{-}reroute}}_{v},
    \label{eq:delta_distance}\\
    \Delta Q_{v}
    &=
    Q^{\mathrm{reroute}}_{v}
    -
    Q^{\mathrm{no\mbox{-}reroute}}_{v},
    \label{eq:delta_co2}
\end{align}
where $S$ denotes trip travel time in simulation steps, $D$ denotes total distance traveled, and $Q$ denotes total $\mathrm{CO}_2$ emissions. Edge-level matched TTI is retained as a diagnostic metric; routes with non-overlapping edges and incomplete trips are retained in post-processing logs where available rather than treated as implicit performance improvements.

Because the simulation step length is one second, $\Delta S_v$ can be interpreted as a trip-level travel-time difference in seconds. Aggregate and percentage-based metrics are computed for CAV-only, non-CAV, and all-vehicle subsets where data are available. The evaluation is repeated across multiple CAV penetration levels to assess scalability and network-wide impacts.

\section{Results and Discussion}\label{Section IV}

To evaluate the proposed collision-aware predictive rerouting framework, simulations were conducted across three urban network scenarios, five CAV penetration levels, three random seeds, and nine routing configurations. The routing configurations include no rerouting, live-state routing without collision-window information, collision-window routing without DCRNN forecasting, DCRNN routing without collision-window information, the full DCRNN collision-window framework, K-shortest path routing, predictive density routing, V2X proactive routing, and SUMO-based rerouting. The completed simulation dataset contains 405 trip-metric output files and 280{,}935 matched trip records. Performance is evaluated using trip-level travel time, distance traveled, total $\mathrm{CO}_2$ emissions, rerouting frequency, and incident-exposure time.

The results are interpreted as a tradeoff analysis rather than as evidence that a single method dominates all criteria. In particular, the full DCRNN collision-window framework does not achieve the lowest aggregate travel time, distance, or emissions across all methods. The strongest consistent advantage of the proposed framework is its much lower rerouting intensity relative to active rerouting baselines, while maintaining similar travel-time and emissions performance. This behavior is important because frequent rerouting can increase route instability, create driver or passenger discomfort, and introduce unnecessary control actions in connected-vehicle deployments.

\begin{table*}[t]
\centering
\caption{Aggregate Trip-Level Performance Across All Evaluated Scenarios}
\label{tab:raw_metrics}
\footnotesize
\resizebox{\textwidth}{!}{%
\begin{tabular}{@{}lccccc@{}}
\toprule
Method & Travel time (s) & Distance (m) & CO$_2$ (mg) & Reroutes/trip & Incident exposure (s) \\
\midrule
No reroute & 231.904 & 1250.502 & 404031.403 & 0.000 & 0.000 \\
Live state, no collision window & 231.257 & 1248.525 & 402857.110 & 0.016 & 0.000 \\
Collision window, no DCRNN & 231.319 & 1248.487 & 403055.586 & 0.016 & 9.616 \\
DCRNN only, no collision window & 231.525 & 1248.821 & 403322.671 & 0.013 & 0.000 \\
Full DCRNN collision-window framework & 231.812 & 1249.019 & 403919.801 & 0.014 & 9.600 \\
K-shortest path & 230.527 & 1244.671 & 401739.198 & 0.030 & 6.970 \\
Predictive density & 230.611 & 1244.691 & 401892.554 & 0.031 & 6.978 \\
V2X proactive & 230.527 & 1244.671 & 401739.198 & 0.030 & 6.970 \\
SUMO-based reroute & 230.700 & 1244.866 & 402106.628 & 0.031 & 9.601 \\
\bottomrule
\end{tabular}%
}
\end{table*}

Table~\ref{tab:raw_metrics} summarizes the simulation-campaign aggregate trip-level metrics. The K-shortest path and V2X proactive baselines achieve the lowest overall mean travel time and distance, and they also produce the lowest aggregate $\mathrm{CO}_2$ emissions. The full DCRNN collision-window framework is therefore not the best-performing method on these aggregate efficiency metrics. However, the differences are small in magnitude. Compared with K-shortest path and V2X proactive routing, the full framework increases mean travel time by approximately 0.56\% and mean $\mathrm{CO}_2$ emissions by approximately 0.54\%. Compared with the no-rerouting baseline, the full framework slightly reduces travel time, distance, and emissions.

The most pronounced distinction is rerouting frequency. The full DCRNN collision-window framework averages 0.014 reroutes per trip, whereas K-shortest path and V2X proactive routing average 0.030 reroutes per trip, predictive density averages 0.031 reroutes per trip, and SUMO-based rerouting averages 0.031 reroutes per trip. Thus, the proposed full framework uses roughly 55--56\% fewer reroutes than these active routing baselines while maintaining similar aggregate travel-time and emissions levels. This result supports a stability-oriented interpretation: the proposed controller is more conservative and generates fewer route changes, rather than aggressively optimizing the shortest aggregate travel-time objective.

The incident-exposure metric requires careful interpretation. Several methods report zero incident exposure even though they are evaluated under the same incident-enabled simulation family. This indicates that incident exposure may depend on method-specific route tracking or exposure-accounting logic and is not used as the sole basis for a broad safety-dominance claim. Accordingly, the results are interpreted as evidence of collision-aware routing behavior and reduced rerouting intensity, but not as proof that the full method universally minimizes incident exposure.

\begin{figure}[t]
\centering
\includegraphics[width=\columnwidth]{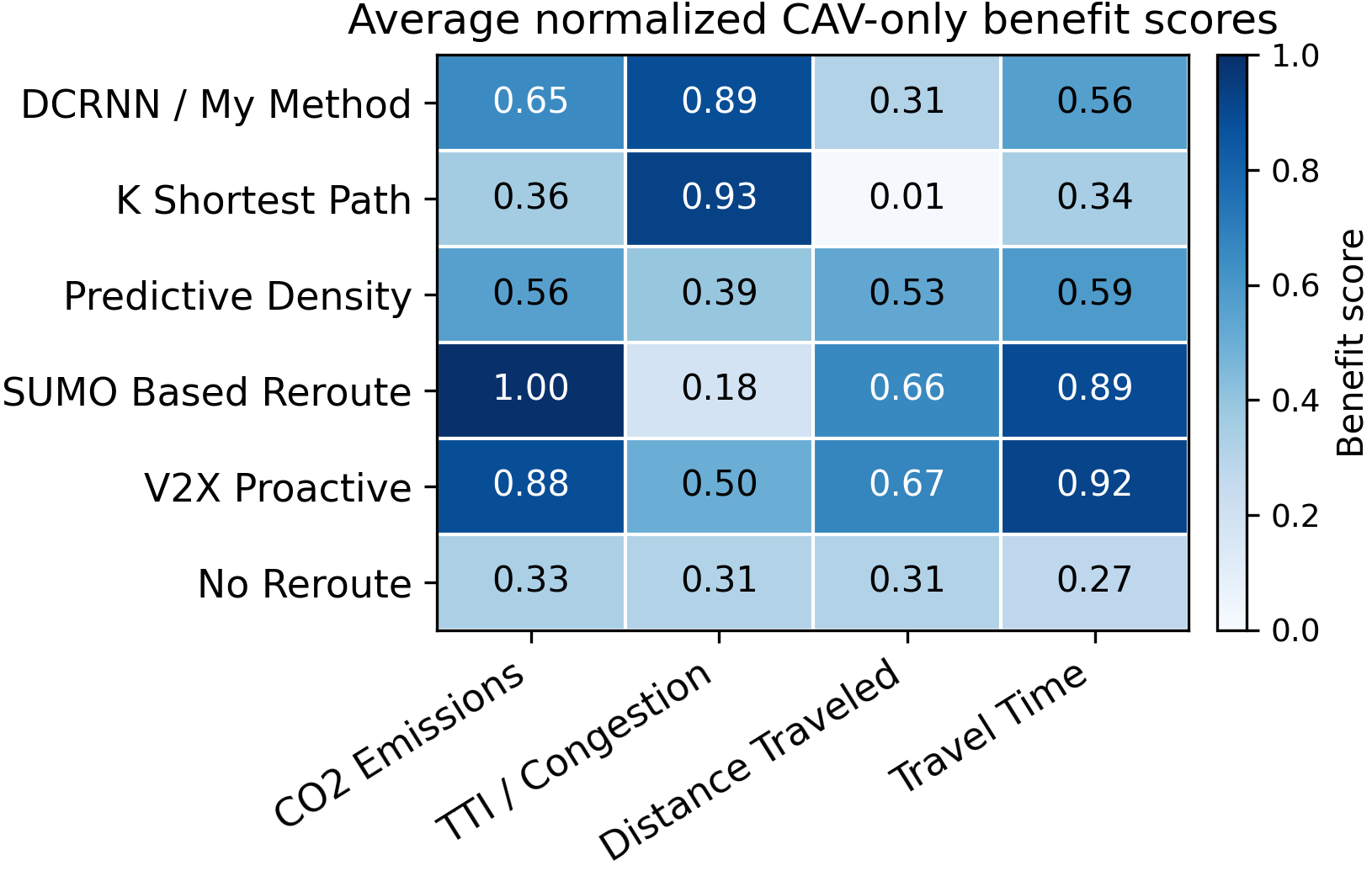}
\caption{Average normalized CAV-only benefit scores for emissions, TTI/congestion, distance traveled, and travel time across routing methods. Scores are normalized within each network and averaged across evaluated scenarios. Higher values indicate stronger relative performance for the corresponding metric.}
\label{fig:benefit_heatmap}
\end{figure}

\subsection{Metric-Level Performance Analysis}

Figure~\ref{fig:benefit_heatmap} presents normalized CAV-only benefit-score comparisons across emissions, TTI/congestion, distance traveled, and travel time. The visualization indicates that the routing methods occupy distinct metric-specific operating profiles rather than following a single dominant ranking. The proposed DCRNN framework and K-shortest path routing obtain the strongest TTI/congestion scores, while SUMO-based rerouting performs strongly on emissions, distance, and travel time but obtains a lower normalized TTI/congestion score. V2X proactive routing also shows strong emissions, distance, and travel-time performance.

These score patterns are consistent with the broader tradeoff interpretation of the simulation campaign. The full DCRNN collision-window framework is not the best-performing method on every diagnostic metric, but it provides a favorable congestion-related score while maintaining a conservative rerouting profile. In contrast, several active routing baselines achieve stronger distance or travel-time scores but apply substantially more route changes, as shown in Table~\ref{tab:raw_metrics}.

The key conclusion from the metric-level analysis is therefore not that the proposed framework dominates all criteria, but that it provides a competitive congestion-oriented operating point while supporting the stability-oriented interpretation of the aggregate trip metrics. The trip-level means in Table~\ref{tab:raw_metrics} remain the primary numerical basis for the main performance claims, while Fig.~\ref{fig:benefit_heatmap} provides a compact diagnostic visualization of metric-specific tradeoffs.

\subsection{Impact of CAV Penetration at Different Percentages}

To examine how routing behavior changes with CAV participation, the simulation-campaign results were aggregated across 5\%, 15\%, 25\%, 35\%, and 50\% CAV penetration levels. Table~\ref{tab:penetration_tradeoff} compares the full DCRNN collision-window framework with the average of four active routing baselines: K-shortest path, predictive density, V2X proactive routing, and SUMO-based rerouting.

\begin{table*}[t]
\centering
\caption{Full DCRNN Collision-Window Framework Relative to Active Routing Baselines by CAV Penetration}
\label{tab:penetration_tradeoff}
\footnotesize
\resizebox{\textwidth}{!}{%
\begin{tabular}{@{}lccc@{}}
\toprule
CAV penetration & Travel time relative to active-baseline average & CO$_2$ relative to active-baseline average & Reroutes relative to active-baseline average \\
\midrule
5\%  & $-0.552\%$ & $+0.051\%$ & $-96.247\%$ \\
15\% & $-1.096\%$ & $-0.650\%$ & $-89.712\%$ \\
25\% & $-0.997\%$ & $-0.579\%$ & $-82.048\%$ \\
35\% & $-0.778\%$ & $-0.494\%$ & $-75.749\%$ \\
50\% & $-0.495\%$ & $-0.354\%$ & $-71.645\%$ \\
\bottomrule
\end{tabular}%
}
\end{table*}

Across penetration levels, the full framework achieves travel times that are approximately 0.50--1.10\% lower than the average of the active routing baselines and $\mathrm{CO}_2$ emissions that are approximately 0.35--0.65\% lower at penetration levels of 15--50\%. At 5\% penetration, the emissions difference is nearly neutral. The largest and most consistent effect is again rerouting frequency: the full framework reduces rerouting by approximately 72--96\% relative to the active-baseline average. This confirms that the proposed controller's main benefit is not aggressive global travel-time minimization, but low-reroute operation with comparable efficiency.

\begin{figure}[t]
\centering
\includegraphics[width=\columnwidth]{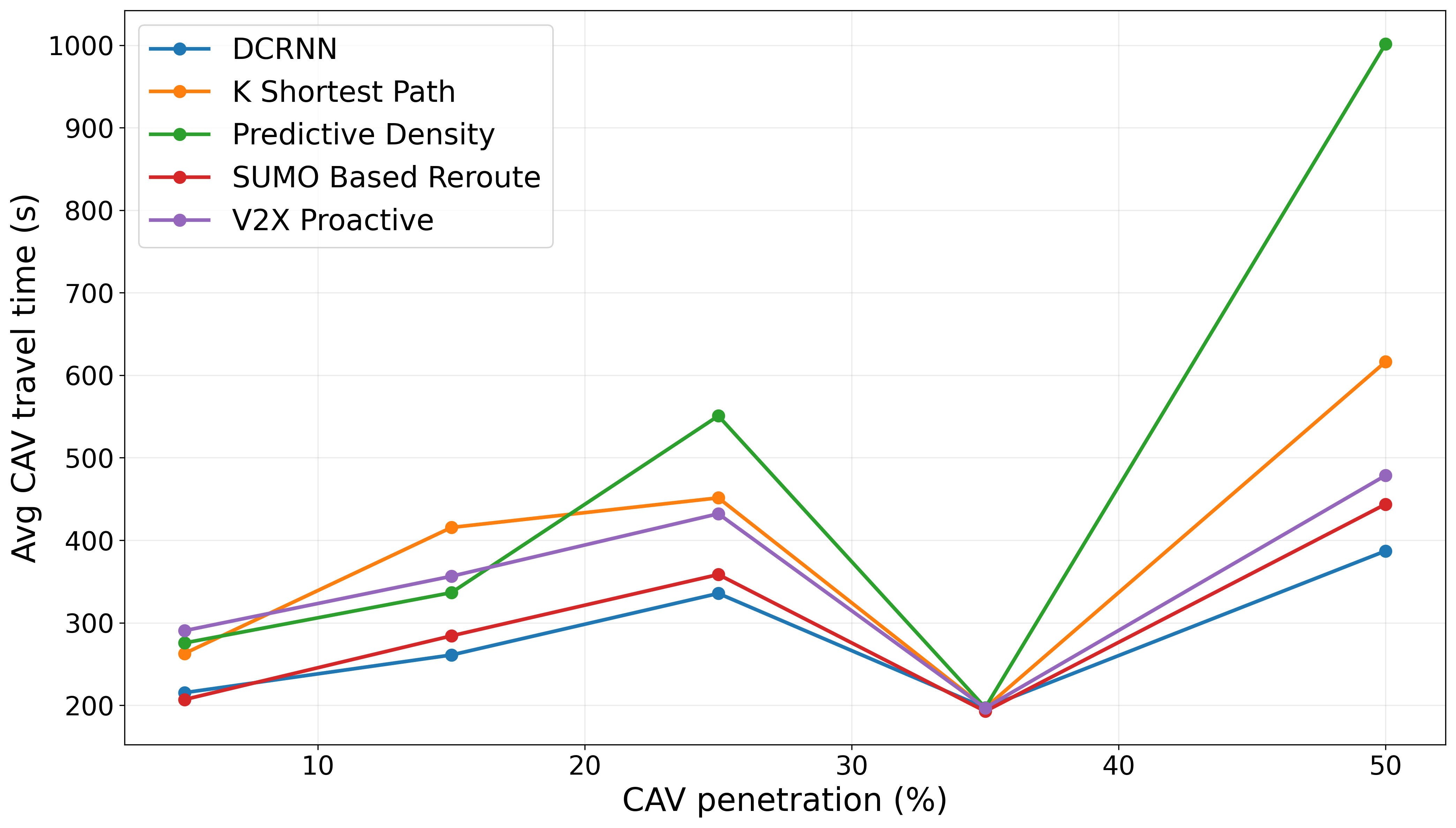}
\caption{Average CAV travel time across penetration levels for the evaluated routing methods.}
\label{fig:cav_trend_travel_time}
\end{figure}

Figure~\ref{fig:cav_trend_travel_time} illustrates how travel-time performance changes with penetration level in the CAV-only trend visualization. The simulation-campaign analysis indicates that penetration-level effects are best interpreted as relative tradeoffs rather than as unconditional dominance. In particular, the full framework can remain competitive with the active-baseline average while applying far fewer route changes.

\begin{figure}[t]
\centering
\includegraphics[width=\columnwidth]{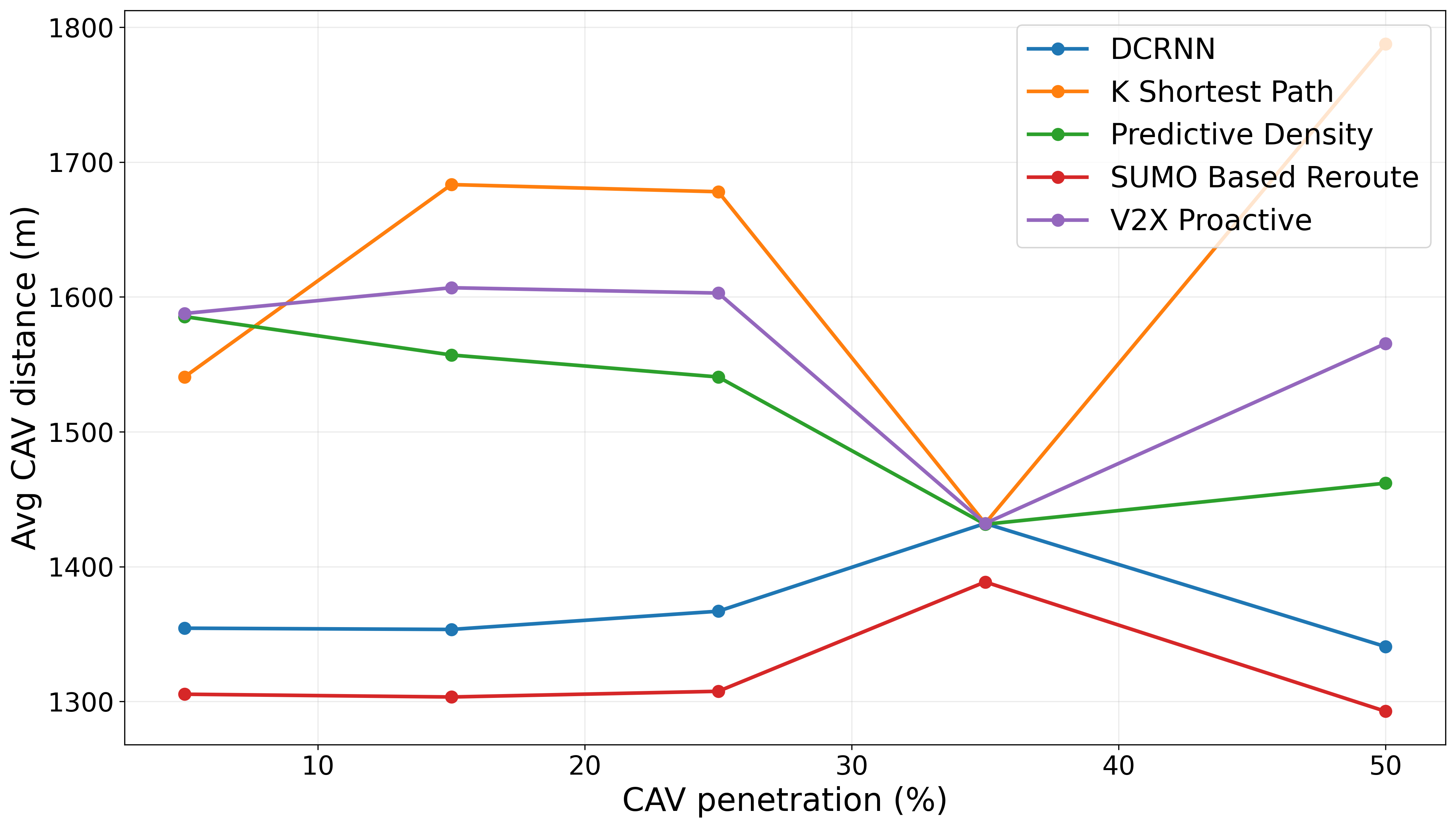}
\caption{Average CAV distance traveled across penetration levels for the evaluated routing methods.}
\label{fig:cav_trend_distance}
\end{figure}

Figure~\ref{fig:cav_trend_distance} shows the corresponding distance trends. Distance is a useful indicator of detour behavior. The aggregate simulation-campaign results show that the most travel-time-efficient active baselines also produce slightly shorter mean distance than the full framework, while the no-rerouting baseline produces the longest mean distance among the reported aggregate metrics.

\begin{figure}[t]
\centering
\includegraphics[width=\columnwidth]{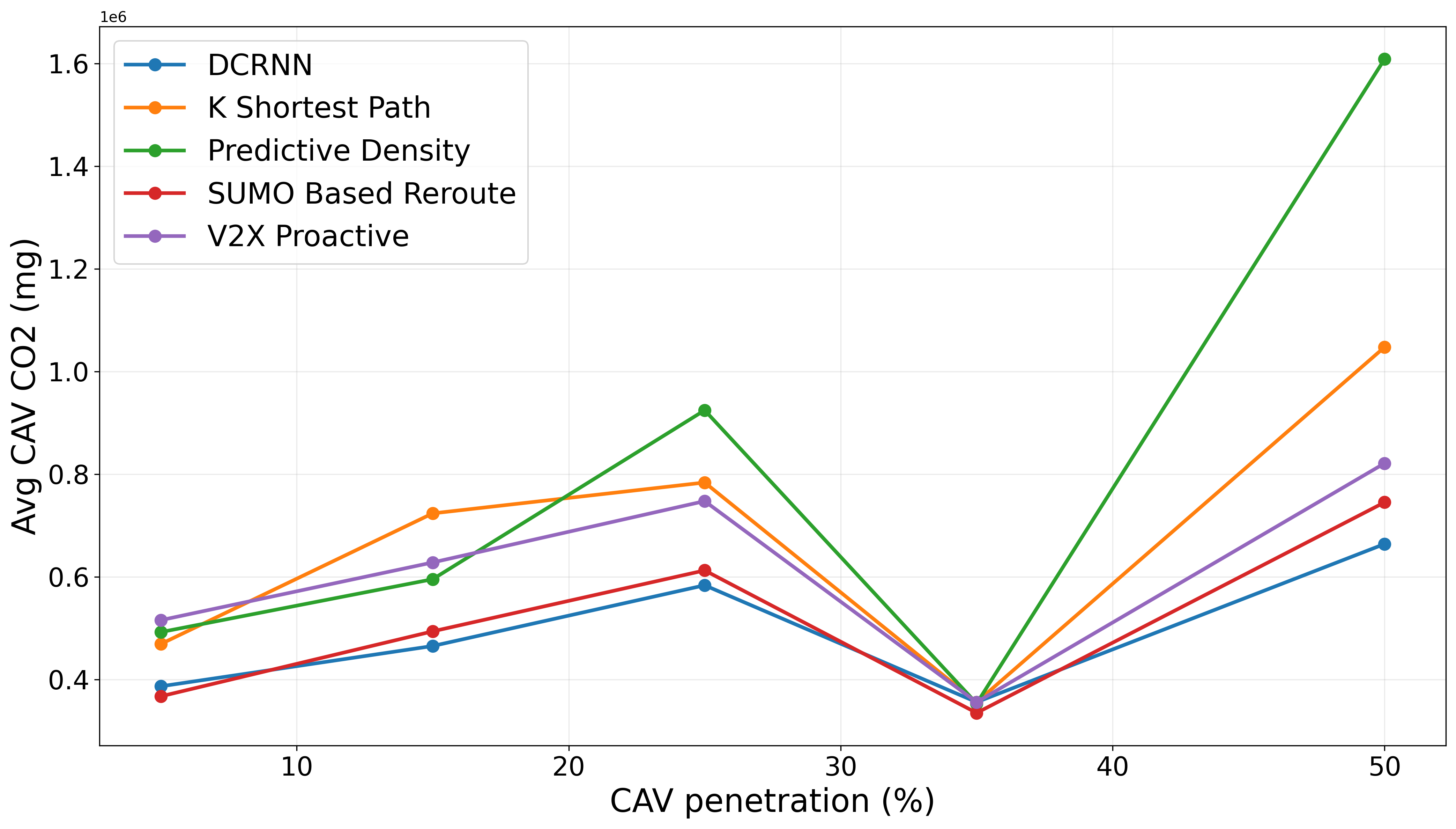}
\caption{Average CAV CO$_2$ emissions across penetration levels for the evaluated routing methods.}
\label{fig:cav_trend_co2}
\end{figure}

Figure~\ref{fig:cav_trend_co2} presents the emissions trend. The simulation-campaign aggregate results indicate that emissions differences among the active strategies are small, generally below one percent. Therefore, the emissions result is described as comparable performance rather than as a large environmental improvement by the proposed method.

\begin{figure}[t]
\centering
\includegraphics[width=\columnwidth]{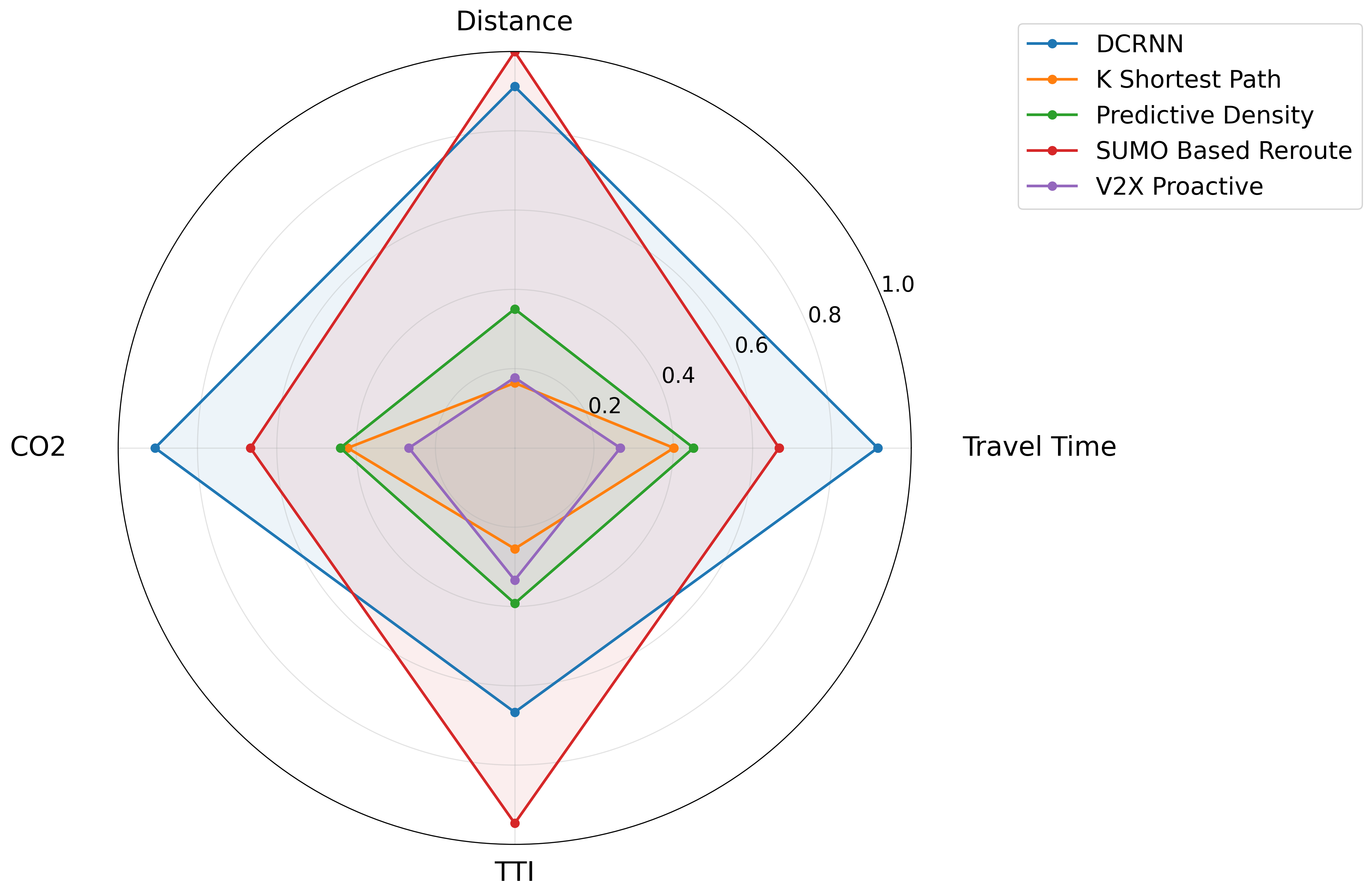}
\caption{Normalized CAV benefit score by routing method at 5\% CAV penetration.}
\label{fig:cav_radar_5_percent}
\end{figure}

The radar visualization at 5\% penetration in Figure~\ref{fig:cav_radar_5_percent} provides an additional multi-objective view of the low-penetration case. At low CAV participation, relatively few vehicles can respond to control recommendations, so large network-wide changes are not expected. The penetration-level tradeoff table shows that the full framework's strongest low-penetration effect is a large reduction in rerouting frequency, with nearly neutral emissions relative to the active-baseline average.

\begin{figure*}[t]
\centering
\includegraphics[width=0.95\textwidth,trim=0 0.08in 0 0.06in,clip]{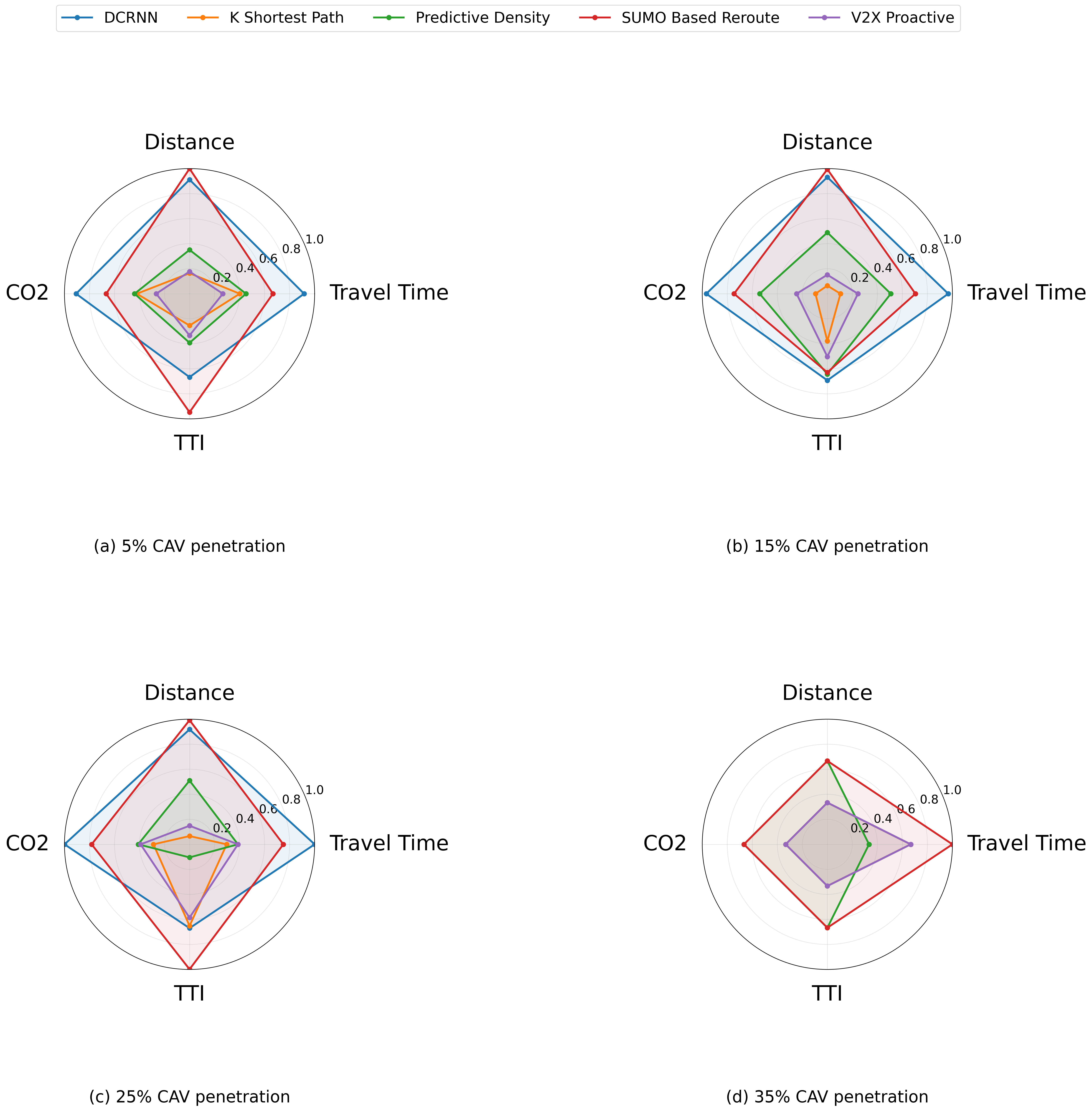}
\vspace{-1.0ex}
\caption{Normalized CAV benefit scores by routing method for the available four-panel radar summary. The panels correspond to the CAV penetration levels shown in the subfigure labels. The 50\% penetration case is summarized quantitatively in Table~\ref{tab:penetration_tradeoff} and in the trend plots.}
\label{fig:cav_radar_penetration_combined}
\end{figure*}

Figure~\ref{fig:cav_radar_penetration_combined} summarizes the available four-panel radar comparison for the penetration levels indicated in the subfigure labels. These radar plots are retained to show how multi-objective performance profiles vary with penetration level. The 50\% penetration case is evaluated using the trend plots and the penetration-level tradeoff results in Table~\ref{tab:penetration_tradeoff}. The revised interpretation is that the proposed framework provides a conservative and stable control profile across penetration levels, while the relative advantage of each benchmark depends on the chosen metric and normalization.

\subsection{Overall Routing Effectiveness}

The overall routing effectiveness analysis is revised to avoid claiming that the proposed method is best overall. The simulation-campaign trip-level metrics show that K-shortest path and V2X proactive routing achieve the lowest aggregate travel time, distance, and emissions. However, these active baselines also apply substantially more reroutes. The full DCRNN collision-window framework therefore occupies a different operating point: it sacrifices a small amount of aggregate efficiency relative to the strongest active baselines while sharply reducing rerouting intensity.

\begin{figure*}[t]
\centering
\includegraphics[width=0.75\textwidth]{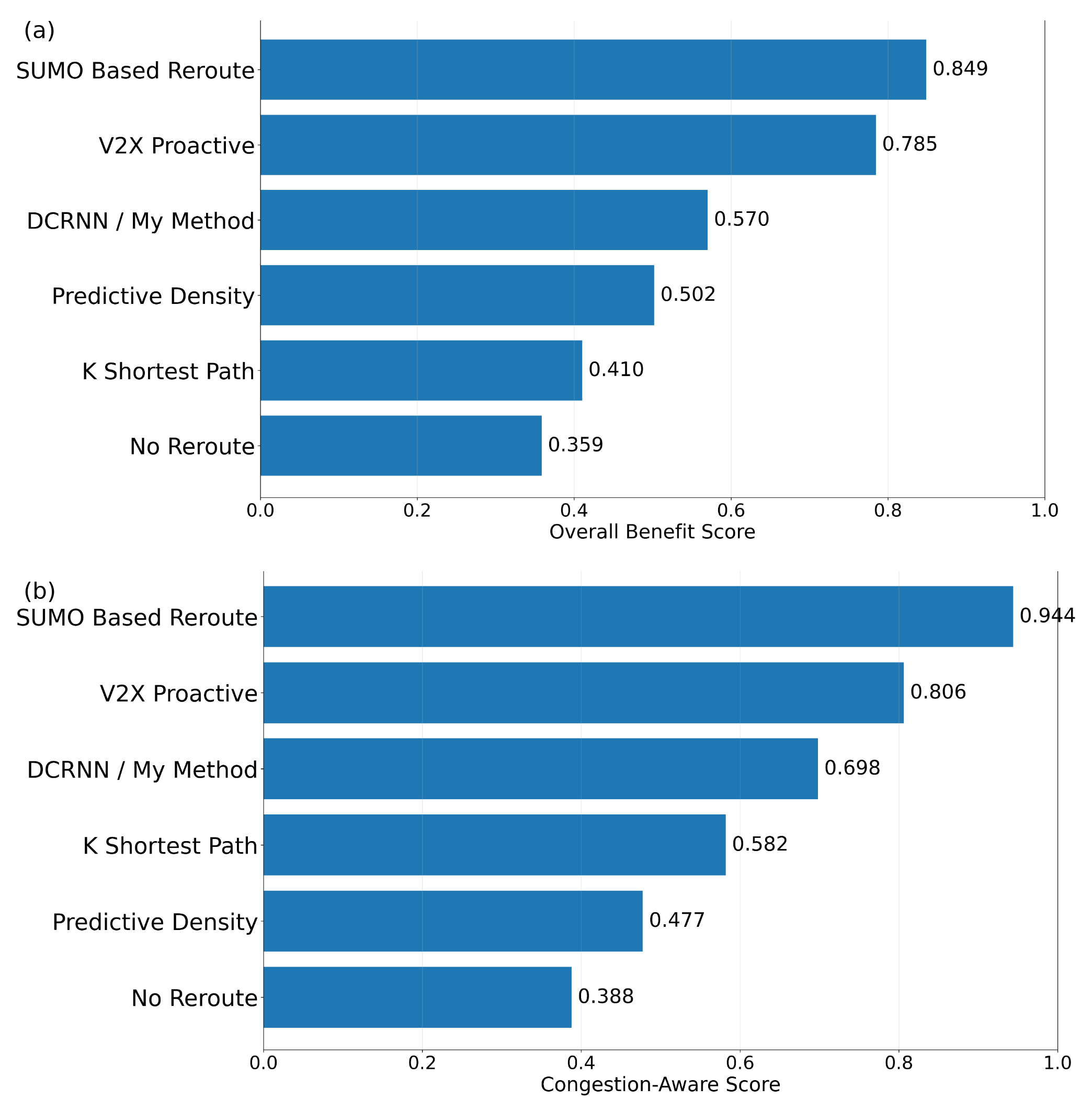}
\caption{Routing performance score comparison: (a) overall routing benefit score computed as the mean of normalized emissions, congestion, distance, and travel-time benefit scores; and (b) congestion-aware routing score computed using weighted contributions from congestion reduction, emissions reduction, and travel-time improvement.}
\label{fig:overall_and_congestion_scores}
\end{figure*}

Figure~\ref{fig:overall_and_congestion_scores} is therefore interpreted as a score-based visualization of selected normalized criteria, not as the sole basis for performance claims. Score-based figures are sensitive to metric selection, weighting, and normalization. For this reason, the discussion emphasizes the simulation-campaign trip-level means and penetration-level tradeoff table, while using the score figure to illustrate the multi-objective nature of the comparison.

\subsection{DCRNN Forecast Validation}

The DCRNN forecast validation does not support a claim that the DCRNN midpoint prediction outperforms all simple forecasting baselines. In the simulation-campaign forecast evaluation, persistence achieves the lowest MAE at the 5~s and 10~s horizons for both DCRNN-only and full DCRNN collision-window methods. At the 30~s horizon, persistence gives the lowest MAE, while the historical-average baseline gives the lowest RMSE. Table~\ref{tab:forecast_validation_summary} summarizes this result.

\begin{table}[t]
\centering
\caption{Best Forecast Baselines in Simulation-Campaign Forecast Validation}
\label{tab:forecast_validation_summary}
\scriptsize
\renewcommand{\arraystretch}{0.95}
\resizebox{0.92\columnwidth}{!}{%
\begin{tabular}{@{}lcc@{}}
\toprule
Horizon & Best MAE baseline & Best RMSE baseline \\
\midrule
5~s  & Persistence & Persistence \\
10~s & Persistence & Persistence \\
30~s & Persistence & Historical average \\
\bottomrule
\end{tabular}%
}
\end{table}

This finding changes the interpretation of the learning component. In the present experimental configuration, the DCRNN is interpreted as a structured spatiotemporal input to the routing framework rather than as a uniformly superior standalone forecasting model. The result motivates future work on retraining, horizon-specific calibration, uncertainty-aware prediction, and direct optimization of the forecast signal for route-choice decisions.

\subsection{Tradeoff Characteristics of Routing Strategies}

\begin{figure}[t]
\centering
\includegraphics[width=\columnwidth]{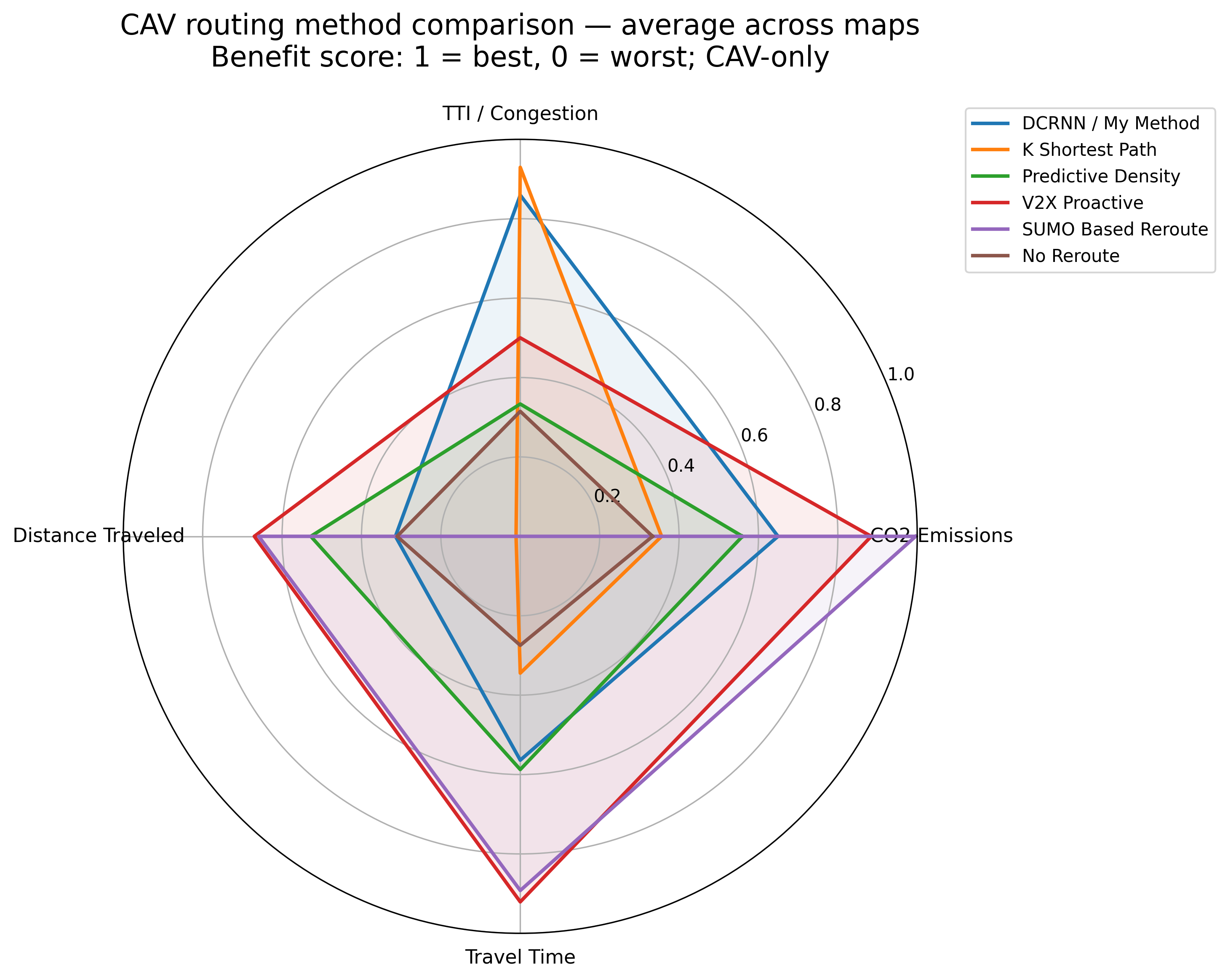}
\caption{Radar-chart comparison of normalized routing benefit scores across evaluation metrics. Larger enclosed areas indicate higher normalized scores under the selected metric scaling, and should be interpreted as a diagnostic visualization of routing tradeoffs rather than as a standalone ranking.}
\label{fig:overall_radar}
\end{figure}

Figure~\ref{fig:overall_radar} provides a comprehensive visualization of routing tradeoffs. The revised analysis indicates that the routing strategies form distinct operating profiles rather than a single strict ranking. K-shortest path and V2X proactive routing are strong aggregate-efficiency baselines. Predictive density and SUMO-based rerouting provide additional reference points for live-state and simulation-based route adaptation. The full DCRNN collision-window framework is most clearly distinguished by its conservative low-reroute behavior.

These results suggest that predictive collision-aware routing can be useful when route stability and control conservatism are important deployment objectives. However, the current results do not justify a broad claim that the proposed method dominates all baselines in travel time, distance, emissions, incident exposure, or forecasting accuracy. The appropriate conclusion is that the proposed method achieves comparable efficiency with much lower rerouting frequency, while highlighting important tradeoffs for additional ablation, sensitivity, and all-vehicle analyses.

\subsection{Limitations}

Several limitations are considered when interpreting the results. First, the implemented collision-window logic assumes that the timing and location of programmed incident events are available to the controller. This is useful for controlled simulation but represents an idealized form of incident awareness; future work will compare exact incident-window information with delayed, uncertain, or partially observed incident reports. Second, the incident scenarios are generated using programmed stops and therefore serve as repeatable surrogates for non-recurrent congestion rather than full physical crash reconstructions. Third, the analysis is simulation-based, and transfer to real-world deployment would require calibration of driver behavior, CAV compliance, sensing latency, communication reliability, and demand uncertainty.

The incident-exposure metric also requires further validation. Some baseline configurations report zero exposure under the current post-processing logic, which suggests that exposure accounting may not be directly comparable across all methods. Future work will use a unified exposure definition applied independently of controller type and will report exposure for CAV, non-CAV, and all-vehicle groups. The current forecast-validation results also show that simple persistence and historical-average baselines can outperform the DCRNN midpoint forecast on aggregate error metrics, indicating a need for forecast recalibration and uncertainty-aware prediction. Finally, additional sensitivity analysis will be conducted for route weights, detour caps, lookahead horizons, improvement thresholds, cooldown periods, reroute limits, and other gate parameters.

\section{Conclusions}\label{Section V}
In this paper, a collision-aware predictive rerouting framework was developed for connected and automated vehicles operating under non-recurrent congestion and incident-induced traffic disruptions. The framework integrates microscopic traffic simulation, DCRNN-based spatiotemporal traffic-state forecasting, collision-window lookahead, and multi-objective route scoring to support real-time adaptive rerouting decisions. The completed simulation campaign across three urban network scenarios, five CAV penetration levels, three random seeds, and nine routing configurations shows that the proposed full framework does not dominate all baselines in aggregate travel time, distance, or emissions. Instead, the strongest result is a stability-oriented tradeoff: the full framework maintains comparable efficiency while substantially reducing rerouting frequency relative to active rerouting baselines. The results indicate that predictive collision-aware routing is best interpreted as a conservative control strategy rather than as an aggressive shortest-time optimizer. Across penetration levels, the full framework reduced rerouting by approximately 72--96\% relative to the active-baseline average, while keeping travel-time and emission differences within about one percent. At the same time, the incident-exposure and forecast-validation results show that further work is needed before making broader safety or forecasting-dominance claims. 

Future work will validate the framework on larger, more diverse networks and develop cooperative multi-vehicle strategies to reduce congestion while preserving fairness and route stability.


\bibliographystyle{IEEEtran}
\bibliography{myBib}

\begin{IEEEbiography}[{\includegraphics[width=1in,height=1.25in,clip,keepaspectratio]{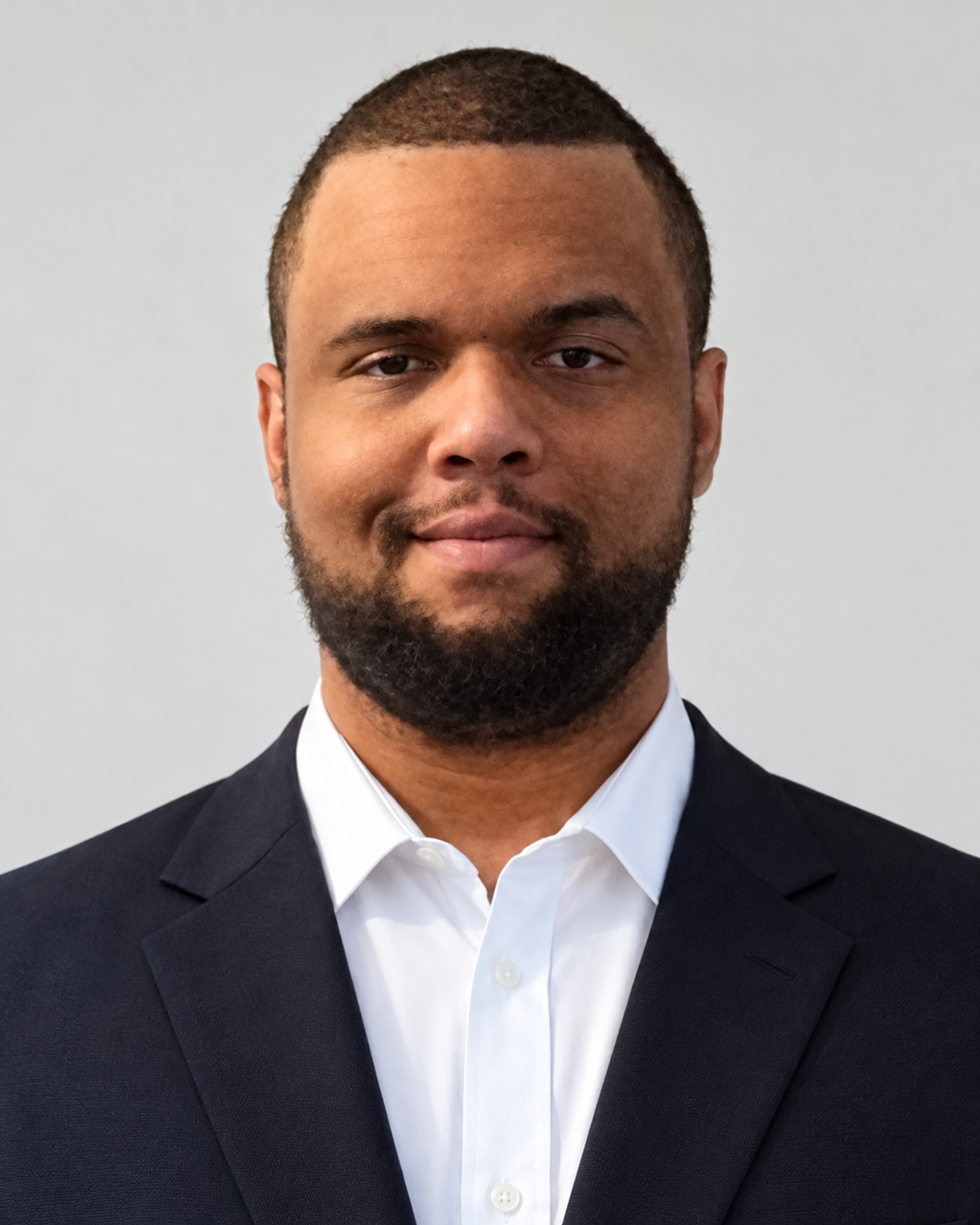}}]{Tony Kinchen} received his B.S. in Biology from Cornell University's College of Agriculture and Life Sciences in 2024. He is currently pursuing the Ph.D. degree in systems engineering at Cornell University, Ithaca, NY, USA, under the supervision of Andreas A. Malikopoulos. His research interests include emerging mobility systems, mobility equity, environmental sustainability, transportation systems, and the development of quantitative metrics for evaluating the relationship between equitable accessibility and environmental impact. His work also explores the use of artificial intelligence and neural networks for traffic anomaly analysis in connected and automated transportation systems.
\end{IEEEbiography}

\begin{IEEEbiography}[{\includegraphics[width=1in,height=1.25in,clip,keepaspectratio]{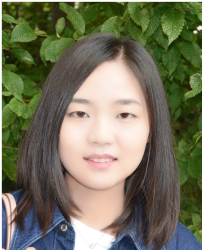}}]{Ting Bai} received the B.Sc. degree in Automation from Northwestern Polytechnical University, Xi'an, China, in 2013, and the Ph.D. degree in Electrical Engineering from Shanghai Jiao Tong University, Shanghai, China, in 2019. From 2020 to 2024, she was a Postdoctoral Researcher at the Department of Electrical Engineering and Computer Science, KTH Royal Institute of Technology, Stockholm, Sweden. From 2024 to 2025, she was a Postdoctoral Research Associate with the School of Civil and Environmental Engineering, Cornell University, Ithaca, NY, USA. Since 2026, she has been an Associate Professor with the School of Automation and Intelligent Sensing, Shanghai Jiao Tong University, Shanghai, China. Her research interests include optimal control, machine learning, reinforcement learning, and their applications to large-scale intelligent transportation systems. Dr. Bai serves as an Associate Editor for IEEE Transactions on Intelligent Transportation Systems.   
\end{IEEEbiography}

\begin{IEEEbiography}[{\includegraphics[width=1in,height=1.25in,clip,keepaspectratio]{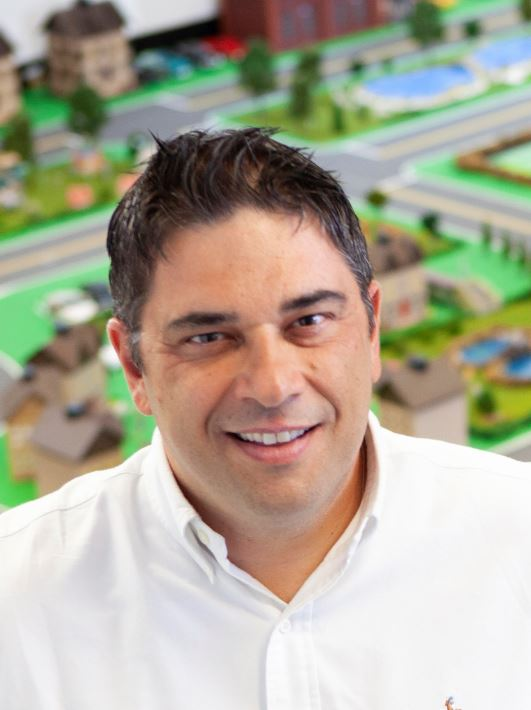}}]{Andreas A. Malikopoulos} (S’06–M’09–SM’17) received the Diploma in mechanical engineering from the National Technical University of Athens, Greece, in 2000, and the M.S. and Ph.D. degrees in mechanical engineering from the University of Michigan, Ann Arbor, MI, USA, in 2004 and 2008, respectively.

He is a Professor in the School of Civil and Environmental Engineering at Cornell University, Ithaca, NY, USA, and the Director of the Information and Decision Science (IDS) Laboratory. He holds courtesy appointments and affiliations with Applied Mathematics, Systems Engineering, Cornell Robotics, Electrical and Computer Engineering, and Mechanical Engineering, and is a Faculty Fellow of the Cornell Atkinson Center for Sustainability. Prior to these appointments, he was the Terri Connor Kelly and John Kelly Career Development Professor in the Department of Mechanical Engineering (2017--2023) and the founding Director of the Sociotechnical Systems Center (2019--2023) at the University of Delaware (UD). Before he joined UD, he was the Alvin M. Weinberg Fellow (2010--2017) in the Energy \& Transportation Science Division at Oak Ridge National Laboratory (ORNL), the Deputy Director of the Urban Dynamics Institute (2014--2017) at ORNL, and a Senior Researcher in General Motors Global Research \& Development (2008--2010). His research interests include stochastic control, team theory, decentralized decision-making, optimization, and the intersection of learning and control, with applications to multi-agent systems, robotics, and emerging mobility systems.

Dr. Malikopoulos received the IEEE Intelligent Transportation Systems (ITS) Society Young Researcher Award in 2019, the IEEE ITS Society Outstanding Application Award in 2026, and Cornell University's Faculty Excellence in Mentoring Postdoctoral Scholars Award in 2026. His research has been recognized with several best paper and student paper awards. He was named a National Academy of Sciences Kavli Frontiers of Science Scholar in 2012. He serves as an Associate Editor of \textit{Automatica} and \textit{IEEE Transactions on Automatic Control}, and as a Senior Editor of \textit{IEEE Transactions on Intelligent Transportation Systems}. He is a member of the IEEE ITS Society Board of Governors and an IEEE ITS Society Distinguished Lecturer for 2024--2027. He is a Fellow of ASME and a member of SIAM, AAAS, and Sigma Xi.
\end{IEEEbiography}

\end{document}